\documentclass[aps,prb,twocolumn]{revtex4}

\newcommand{\appropto}{\mbox{{\raisebox{-0.4ex}{$\stackrel{~\propto~}{{\scriptstyle\sim}}$}}}}

\usepackage{graphicx}
\usepackage{dcolumn}
\usepackage{amsmath}
\usepackage{amssymb}
\usepackage{wasysym}
\usepackage{color}

\usepackage{mathtools}

\begin{document}

\title{Magnetic field-tuned quantum criticality in a Kondo insulator
}

\author{Satya K. Kushwaha,$^{1,2}$ Mun K. Chan$^{1}$, Joonbum Park,$^{1}$ S.~M.~Thomas,$^2$ Eric D. Bauer$^{2}$, J. D. Thompson$^{2}$, F. Ronning$^{2}$, Priscila F. S. Rosa$^{2}$, Neil Harrison$^{1}$}
\affiliation{$^1$MPA-MAG, Los Alamos National Laboratory, Los Alamos NM 87545, USA} 
\affiliation{$^2$MPA-CMMS, Los Alamos National Laboratory, Los Alamos NM 87545, USA} 

\date{\today}

\begin{abstract}
Kondo insulators are predicted to undergo an insulator-to-metal transition under applied magnetic field, yet the extremely high fields required to date have prohibited a comprehensive investigation of the nature of this transition. Here we show that Ce$_3$Bi$_4$Pd$_3$ provides an ideal platform for this investigation, owing to the unusually small magnetic field of $B_{\rm c}\approx$~11~T required to overcome its Kondo insulating gap. Above $B_{\rm c}$, we find a magnetic field-induced Fermi liquid state whose characteristic energy scale $T_{\rm FL}$ collapses near $B_{\rm c}$ in a manner indicative of a magnetic field-tuned quantum critical point. A direct connection is established with the process of Kondo singlet formation, which yields a broad maximum in the magnetic susceptibility as a function of temperature in weak magnetic fields that evolves progressively into a sharper transition at $B_{\rm c}$ as $T\rightarrow$~0. 
\end{abstract}
\maketitle

\section{Introduction}

Kondo insulators are a class of quantum materials in which the coupling between conduction electrons and nearly localized $f$-electrons may lead to properties that are distinct from those of conventional band insulators.\cite{menth1992,aeppli1992,mott1974} Remarkable properties of current interest include topologically protected surface states that are predicted\cite{dzero1,takimoto1,dzero2} and reportedly confirmed by experiments,\cite{kim1,wolgast1,jiang2013,kim2014,li1,zhang2013} and reports of magnetic quantum oscillations originating from the insulating bulk.\cite{tan1,hartstein1,xiang2018,thomas2019} The very same magnetic field that produces quantum oscillations also couples to the $f$-electron magnetic moments, driving the Kondo insulator inexorably towards a metallic state.\cite{cooley1999,sugiyama1988} The required magnetic fields in excess of 100~T\cite{terashima2017,cooley1999} have, however, proven to be prohibitive for a complete characterization of the metallic ground state. 

Thermodynamic experiments have thus far provided evidence for the presence of electronic correlations in Kondo insulators at high magnetic fields. Quantum oscillation experiments, for example, have found moderately heavy masses within the insulating phase in strong magnetic fields.\cite{xiang2018} Furthermore, heat capacity experiments have shown that the electronic contribution undergoes an abrupt increase with increasing magnetic field:\cite{jaime2000,terashima2018} in one case\cite{jaime2000} the increase occurs within the insulating phase suggesting the presence of in-gap states,\cite{boebinger1995} whereas in another, it coincides\cite{terashima2018} with the onset of an upturn in the magnetic susceptibility\cite{sugiyama1988,terashima2017} and reports of metallic behavior.\cite{sugiyama1988,li1} An unambiguous signature of metallic behavior, such as electrical resistivity that increases with increasing temperature at accessible magnetic fields, has yet to be reported. 

\begin{figure}
\includegraphics[angle=0,width=0.98\hsize]{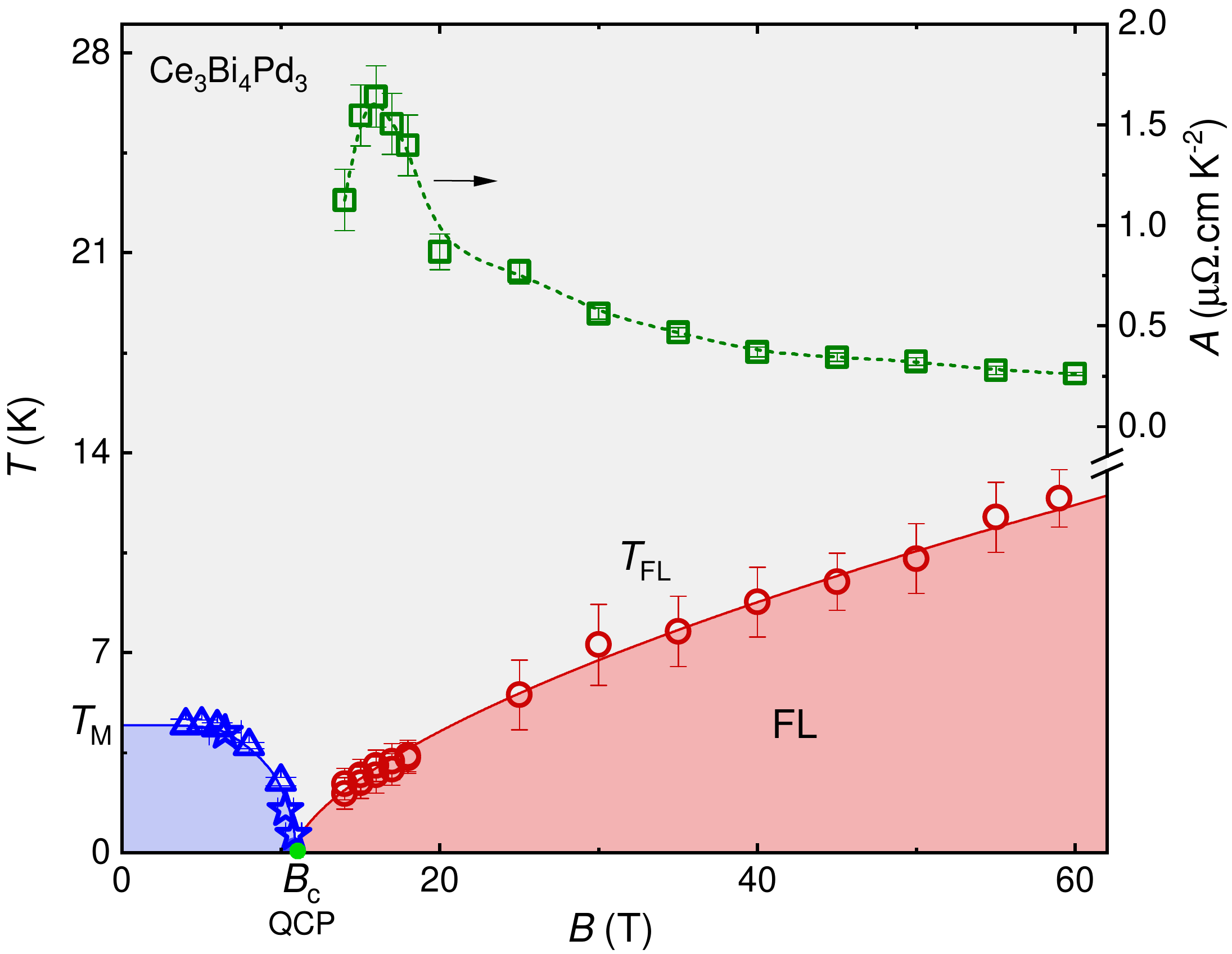}
\caption{Magnetic field versus temperature phase diagram, showing $T_{\rm M}$ obtained from the maxima in $\chi$ versus $T$ (triangles) and $B$ (stars) from Fig.~\ref{MAG}. Here $B_{\rm c}=B(T_{\rm M}\rightarrow$~0). Also shown is the temperature $T_{\rm FL}$ (circles) above which the resistivity departs from $\rho_{xx}=\rho_0+AT^2$ (i.e. Fermi liquid) behavior. Shaded regions and lines are drawn as guides to the eye. The right-hand-axis indicates the magnetic field-dependence of the $A$ coefficient (squares), with the values for $B\lesssim$~20~T corresponding to the average of samples A and B in Fig.~\ref{GAP}.}
\label{diagram}
\end{figure}

We show here that insulating Ce$_3$Bi$_4$Pd$_3$, once  polished to remove surface contamination, exhibits properties consistent with it being a reduced gap variant of its sister topological Kondo insulator candidate, Ce$_3$Bi$_4$Pt$_3$.\cite{boebinger1995,riseborough2000,riseboroughMM,wakeham2016} The atypically small magnetic field of $B_{\rm c}\approx$~11~T required to overcome the Kondo gap of insulating Ce$_3$Bi$_4$Pd$_3$ (see Fig.~\ref{diagram}), makes this material\cite{hermes2008}  ideal for investigating the metallization of a Kondo insulator in strong magnetic fields. We find a magnetic field-induced metallic state exhibiting an electrical resistivity that varies as $T^2$ at low temperatures, thereby revealing a Fermi liquid ground state in the high magnetic field metallic phase. We also identify a magnetic field-tuned collapse of the Fermi liquid temperature scale $T_{\rm FL}$ at $B_{\rm c}$, indicative of a magnetic field-tuned quantum critical point analogous to that observed in SmB$_6$ as a function of pressure.\cite{zhou2017,gabani2003,cooley1995,barla2005,derr2006} 

The origin of the quantum criticality in Ce$_3$Bi$_4$Pd$_3$ is revealed by susceptibility measurements, which find a sharp peak in the susceptibility as a function of magnetic field at $B_{\rm c}$ that can be traced back to a broad maximum in the susceptibility as a function of temperature at $T_{\rm M}$ in weak magnetic fields. This maximum implies the formation of Kondo singlets at a temperature $T_{\rm K}=cT_{\rm M}$ (where 3~$<c<$~4),\cite{jaime2000} yielding 15~$<T_{\rm K}<$~20~K. A gradual evolution of the susceptibility from a crossover along the temperature axis to a sharper transition along the magnetic field axis at $B_{\rm c}$ is therefore unveiled.

\section{Results}

\begin{figure}
\includegraphics[angle=0,width=0.85\hsize]{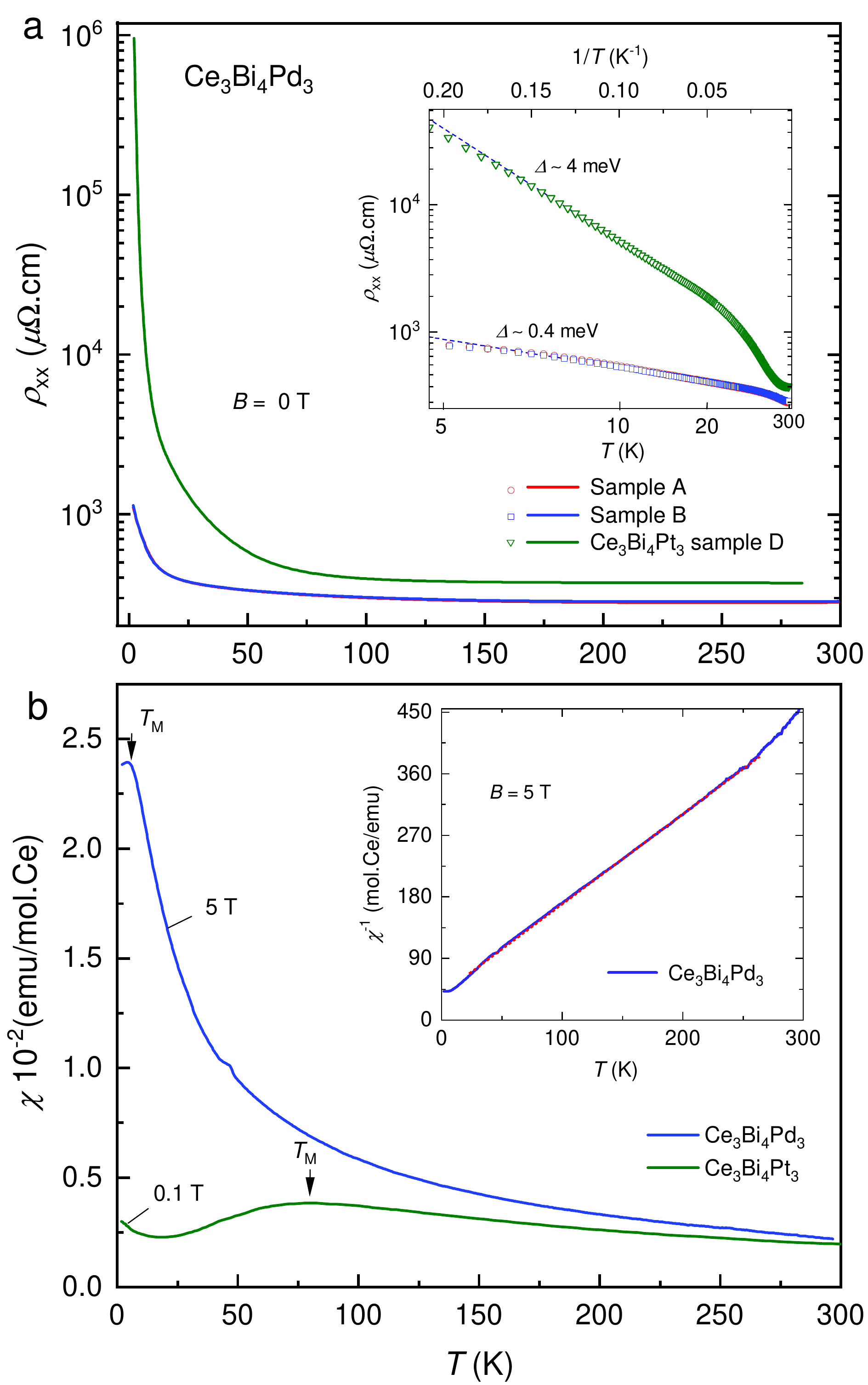}
\caption{{\bf a}, Comparison of $\rho_{xx}$ versus $T$ at $B=0$ for two Ce$_3$Bi$_4$Pd$_3$ samples (labeled A and B) with Ce$_3$Bi$_4$Pt$_3$. The inset compares the logarithmically-scaled electrical resistivity, $\rho_{xx}$, of the two Ce$_3$Bi$_4$Pd$_3$ samples (right-hand-axis) with that of a Ce$_3$Bi$_4$Pt$_3$ sample (left-hand-axis) plotted versus $1/T$ and $T$ with reciprocal scaling. The inset shows an Arrhenius plot. Whereas Ce$_3$Bi$_4$Pt$_3$ exhibits activated ($\rho_{xx}\appropto e^{-\Delta/k_{\rm B}T}$) behavior for 5~$\lesssim$~T~$\lesssim$~30~K with a slope of $\Delta\approx$~4~meV, 
Ce$_3$Bi$_4$Pd$_3$ appears to exhibit such behavior over a range 6~$\lesssim$~T~$\lesssim$~35~K with a slope of $\sim$~0.4~meV ($\sim$~5~K) that lies outside the fitting range, suggesting that $\Delta$ cannot be estimated reliably from $\rho_{xx}$ in the case of Ce$_3$Bi$_4$Pd$_3$. The shallow slope and departure from activated behavior at low $T$ is suggestive of a temperature-dependent scattering rate or residual in-gap states associated either with bulk defects or the sample surface (see Appendix). {\bf b}, Susceptibility $\chi$ versus $T$ for Ce$_3$Bi$_4$Pd$_3$ (at $B=$~5~T) compared with that for Ce$_3$Bi$_4$Pt$_3$ (at $B=$~0.1~T) from Ref.~\cite{hundley1990}, with $T_{\rm M}$ indicating the maxima. The inset shows $1/\chi$ versus $T$ for Ce$_3$Bi$_4$Pd$_3$, evidencing Curie-Weiss behavior (indicated by red dotted fitted line) over a broad range of $T$ and a Curie constant $C\approx$~0.8~mol.Ce/emu~K.}
\label{RES}
\end{figure}

Typical signatures of a Kondo insulating state include an electrical resistivity that exhibits a thermally activated  behavior over a range of temperatures and a Curie-Weiss behavior in the magnetic susceptibility at high temperatures that crosses over via a maximum in the susceptibility into a Kondo gapped state at low temperatures.\cite{boebinger1995,menth1992,aeppli1992,maple1971,kasaya1985,nickerson1971,hundley1990} Electrical resistivity and magnetic susceptibility measurements reveal Ce$_3$Bi$_4$Pd$_3$ to exhibit such signatures (see Fig.~\ref{RES}), but with a Kondo temperature scale that is significantly reduced relative to those found in the archetypal Kondo insulators SmB$_6$ and Ce$_3$Bi$_4$Pt$_3$. Typical Kondo insulators have Kondo gaps of several millielectronvolts, accompanied by inverse residual resistivity ratios in the range 10$^2$~$<R^{-1}_{\rm RRR}<$~10$^5$ and maxima in the magnetic susceptibility and heat capacity\cite{phelan2014,jaime2000} in the range of temperatures 10~K~$\ll T_{\rm M}\lesssim$~100~K. Ce$_3$Bi$_4$Pd$_3$, however, appears to have a significantly smaller Kondo gap, accompanied by an inverse residual resistivity ratio of only $R^{-1}_{\rm RRR}\sim$~5 (see Fig.~\ref{RES}a) and maxima in the susceptibility (see Fig.~\ref{RES}b) and heat capacity (see Appendix) at $T_{\rm M}\approx$~5~K. 

\begin{figure}
\includegraphics[angle=0,width=0.99\hsize]{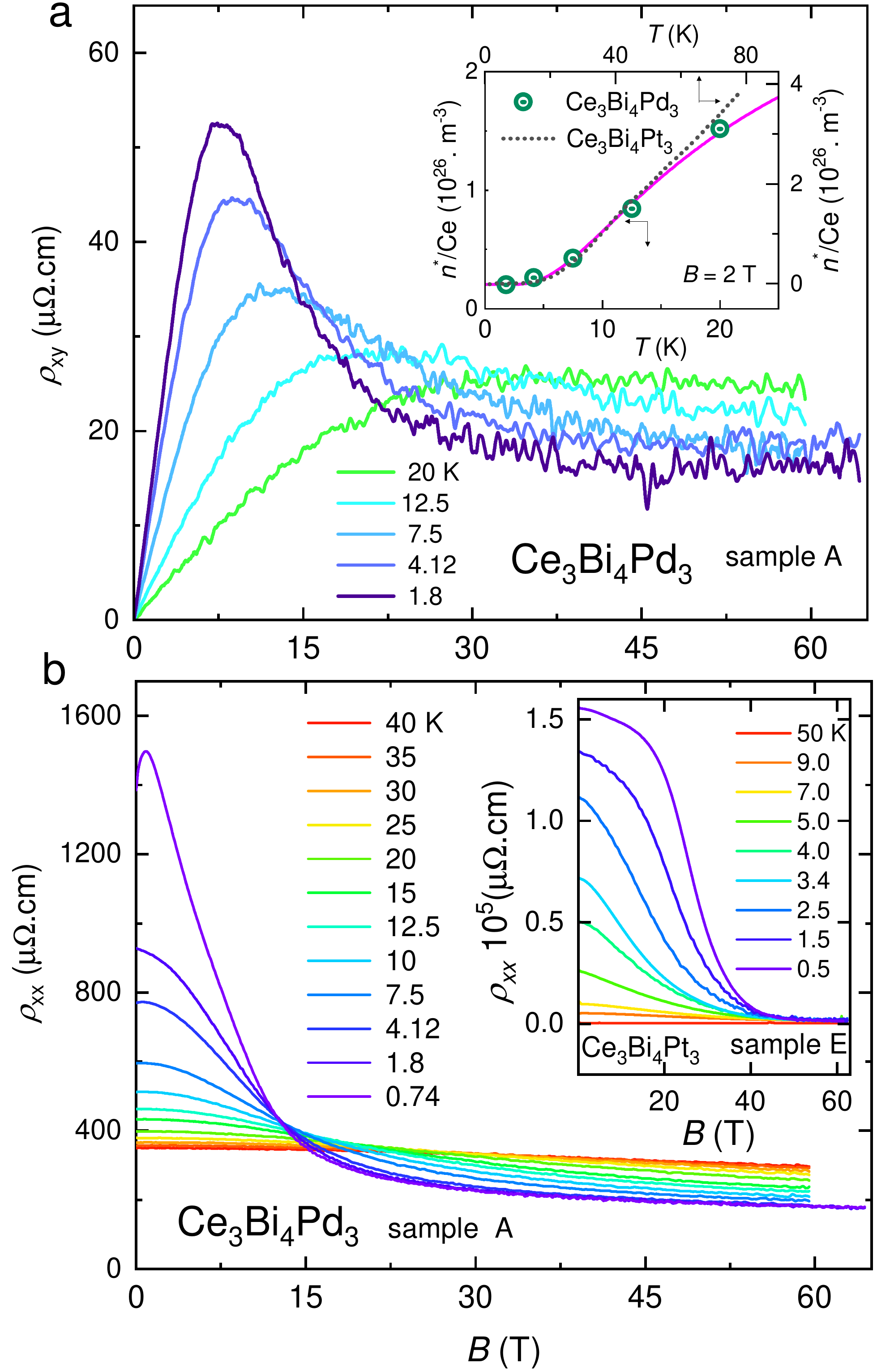}
\caption{{\bf a}, The Hall resistivity $\rho_{xy}$ plotted versus $B$ measured at several temperatures. The inset shows $n^\ast$ (circles) plotted versus $T$ for Ce$_3$Bi$_4$Pd$_3$ at $B=$~2~T (left-hand and bottom axes), estimated using $n^\ast\approx B/e\rho_{xy}$ (not corrected for the anomalous Hall component), with the magenta line representing a fit to $n^\ast=n_0^\ast+n^\ast_1\exp(-\Delta/k_{\rm B}T)$, where $n_0^\ast=$~0.2~$\times10^{26}$~m$^{-3}$, and $n_1^\ast=$~3.7~$\times10^{26}$~m$^{-3}$ and $\Delta=$~1.8~$\pm$~0.5~meV. The finite $n^\ast_0$ for $T\rightarrow0$ is suggestive of residual in-gap states associated either with bulk defects or the sample surface (see Appendix). For comparison, the dotted line shows $n^\ast$ plotted versus $T$ for Ce$_3$Bi$_4$Pt$_3$ at $B=$~5~T (right-hand and top axes) from Ref.~\cite{pietrus2000}. {\bf b}, Measured magnetoresistivity $\rho_{xx}$ of Ce$_{3}$Bi$_{4}$Pd$_3$ sample A (for $B\|\langle100\rangle$) at different temperatures (as indicated in different colors) plotted versus $B$. For comparison, the inset shows the magnetoresistance of Ce$_3$Bi$_4$Pt$_3$ at different temperatures. }
\label{MRHall}
\end{figure}

The small size of the Kondo gap in Ce$_3$Bi$_4$Pd$_3$ in Fig.~\ref{RES}a gives rise to the shallow slope of the resistivity Arrhenius plot in Fig.~\ref{RES}a, which has an increased likelihood of being affected by the presence of in-gap states and changes in the transport scattering rate with temperature. To obtain a more reliable estimate of the Kondo gap, we turn to the activated behavior observed in Hall effect measurements\cite{ashcroft1976} (see Fig.~\ref{MRHall}a), whereupon we obtain $\Delta=$~1.8~$\pm$~0.5~meV ($\approx$~21~K). An important factor in being able to extract the carrier density is that $\rho_{xy}$ (shown in Fig.~\ref{MRHall}a) is linear in $B$ at low $B$. We therefore find that $\Delta\approx k_{\rm B}T_{\rm K}$, which is quantitatively consistent with the hybridized many body band picture.\cite{dzero1} The activated behavior can also be approximately rescaled against that measured in Ce$_3$Bi$_4$Pt$_3$ (see inset of Fig.~\ref{MRHall}a)\cite{pietrus2000} -- the only significant difference being that Ce$_3$Bi$_4$Pd$_3$ has a several times smaller gap and a higher concentration of in-gap states, $n^\ast_0$, than Ce$_3$Bi$_4$Pt$_3$. The atypically small Kondo gap in Ce$_3$Bi$_4$Pd$_3$ implies that this system lies closer to the insulator-metal threshold than other Kondo insulators. Consistent with Ce$_3$Bi$_4$Pd$_3$ having a several times smaller gap, the insulating state is found to be suppressed by magnetic fields that are several times weaker in strength than those required in Ce$_3$Bi$_4$Pt$_3$ (see Fig.~\ref{MRHall}b).



Though a subset of Ce$_3$Bi$_4$Pd$_3$ samples are reported to exhibit a downturn behavior in $\rho_{xx}$ at low temperatures and weak magnetic fields,\cite{dzsaber2017} we focus our attention here only on samples that display an insulating behavior at $B=0$. We consistently prepare such samples (two examples of which are shown in Fig.~\ref{RES}) by polishing away surface oxidation (see Appendix) and by avoiding superconducting binaries inherent to the flux technique used to grow these materials. All samples prepared in such a manner have an electrical resistivity that increases with decreasing $T$ over the entire temperature range probed (see Fig.~\ref{RES}a), in addition to the thermally-activated carrier density at low magnetic fields (see Fig.~\ref{MRHall}a).

\begin{figure}
\includegraphics[angle=-0,width=0.99\hsize]{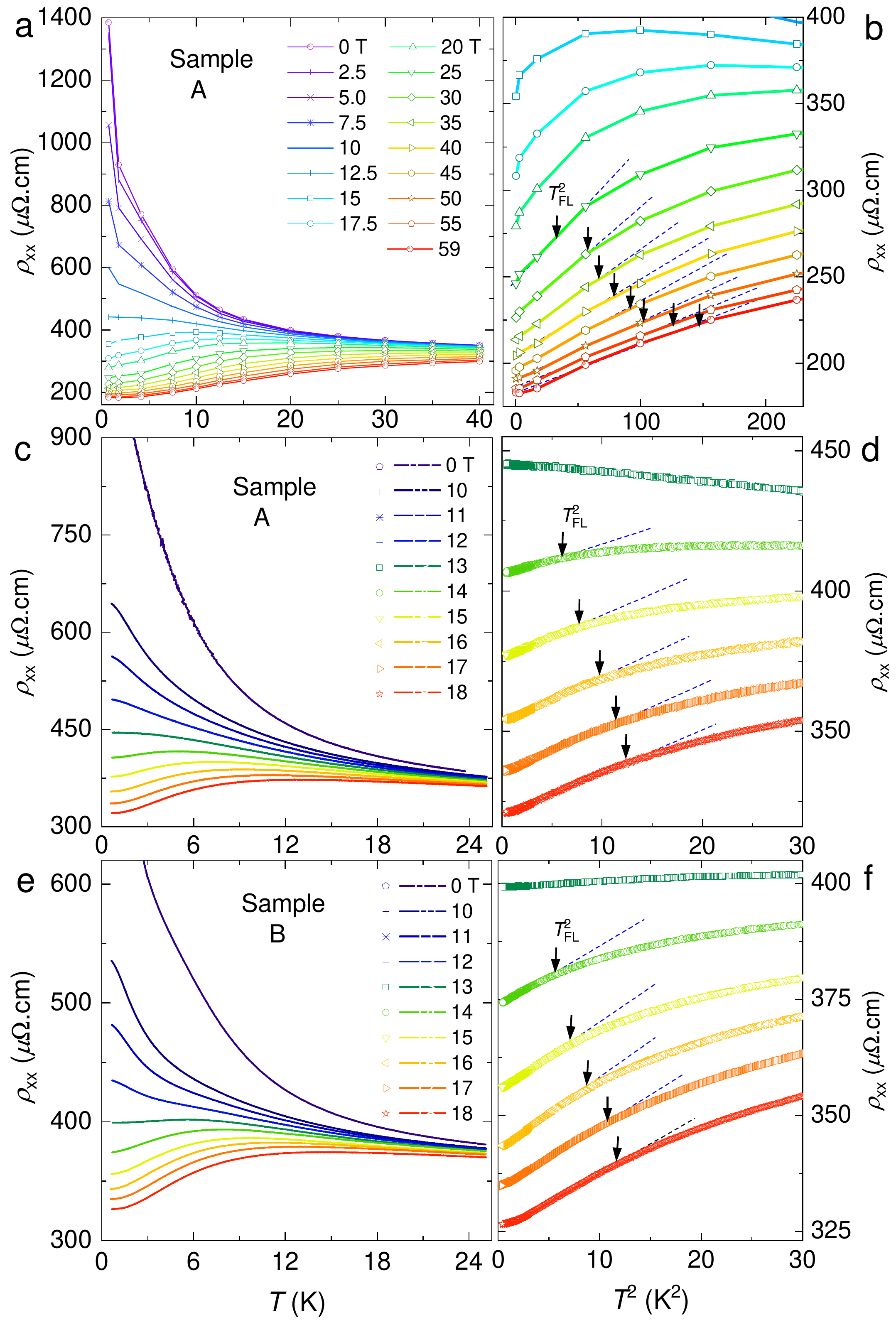}
\caption{{\bf a}, Electrical resistivity $\rho_{xx}$ versus $T$ for sample A at various values of the magnetic field (as indicated) extracted from Fig.~\ref{MRHall}a. Connecting lines are a guide to the eye. {\bf b}, $\rho_{xx}$ versus $T^2$  at various values of the magnetic field (as indicated) extracted from Fig.~\ref{MRHall}a.  {\bf c}, $\rho_{xx}$ versus $T$ for the same sample measured at various steady static magnetic fields (as indicated). {\bf d}, Same $\rho_{xx}$ versus $T^2$ at various steady static magnetic fields (as indicated). Dashed lines indicate the linearity in $T^2$ while arrows indicate $T_{\rm FL}$ (the temperature above which $\rho_{xx}$ deviates from $T^2$ behavior). The slopes of the lines versus $T^\ast$ provide an estimate of $A$ (plotted in Fig.~\ref{diagram}). {\bf e} and {\bf f}, Same as for {\bf c} and {\bf d} but measured on sample B.}
\label{GAP}
\end{figure}

Metallization of the Kondo insulating Ce$_3$Bi$_4$Pd$_3$ samples at $B_{\rm c}\approx$~11~T is evidenced in Figs.~\ref{MRHall}b and \ref{GAP} by the crossover from insulating behavior for $B<B_{\rm c}$ to metallic behavior for $B>B_{\rm c}$. We identify metallic behavior as an electrical resistivity having a value that decreases with decreasing temperature in the limit $T\rightarrow$~0. Further evidence for the transition from insulating to metallic behavior with the closure of the Kondo insulating gap in Ce$_{3}$Bi$_{4}$Pd$_3$ is provided by Hall effect measurements at high magnetic fields. The Hall resistivity $\rho_{xy}$ begins to drop on the approach to $B_{\rm c}$, which indicates an increase in carrier density in advance of the closing of the Kondo gap, likely caused by thermal excitations of carriers across the reduced gap. The nonlinearity of $\rho_{xy}$ above $B_{\rm c}$ indicates the presence of multiple bands and a possible anomalous Hall contribution from skew scattering\cite{fert1987} (see Appendix). In the absence of a dominant anomalous Hall component from skew scattering, the positive Hall coefficient implies either that the hole carriers have numerical supremacy over electron carriers, or that the hole carriers are of higher mobility.\cite{ashcroft1976}

On entering the high magnetic field regime, we find the low-temperature electrical resistivity to display the Fermi liquid form $\rho_{xx}=\rho_0+AT^2$, where $\rho_0$ is a constant and $A$ is a coefficient proportional to the square of the quasiparticle effective mass. The quadratic-in-temperature form is demonstrated by plotting the electrical resistivity versus $T^2$ in Fig.~\ref{GAP}, yielding linear behavior at the lowest temperatures. Consistent with a quantum critical point at $B_{\rm c}$, the Fermi liquid temperature scale $T_{\rm FL}$ is found to collapse on approaching $B_{\rm c}$ in Fig.~\ref{diagram}.\cite{coleman2001} We define $T_{\rm FL}$ as the point beyond which the resistivity is observed to exhibit a downward curvature from $T^2$ behavior on increasing the temperature. At the same time, $A$ is observed to undergo a steep upturn on approaching $B_{\rm c}$ from high magnetic fields in Fig.~\ref{diagram}.  Both the upturn in $A$ and collapse of $T_{\rm FL}$ are indicative of a collapsing Fermi liquid energy scale in the vicinity of $B_{\rm c}$, which is one of the primary signatures of quantum criticality in Fermi liquid systems.\cite{coleman2001} Rather than peaking precisely at $B_{\rm c}$, $A$ is observed to peak at $B\approx$~15~T; a likely consequence of the metallic contribution to the conductivity vanishing due to the collapse of the sizes of the Fermi surface pockets in the limit $B\rightarrow B_{\rm c}$.

\begin{figure}
\includegraphics[angle=0,width=0.9\hsize]{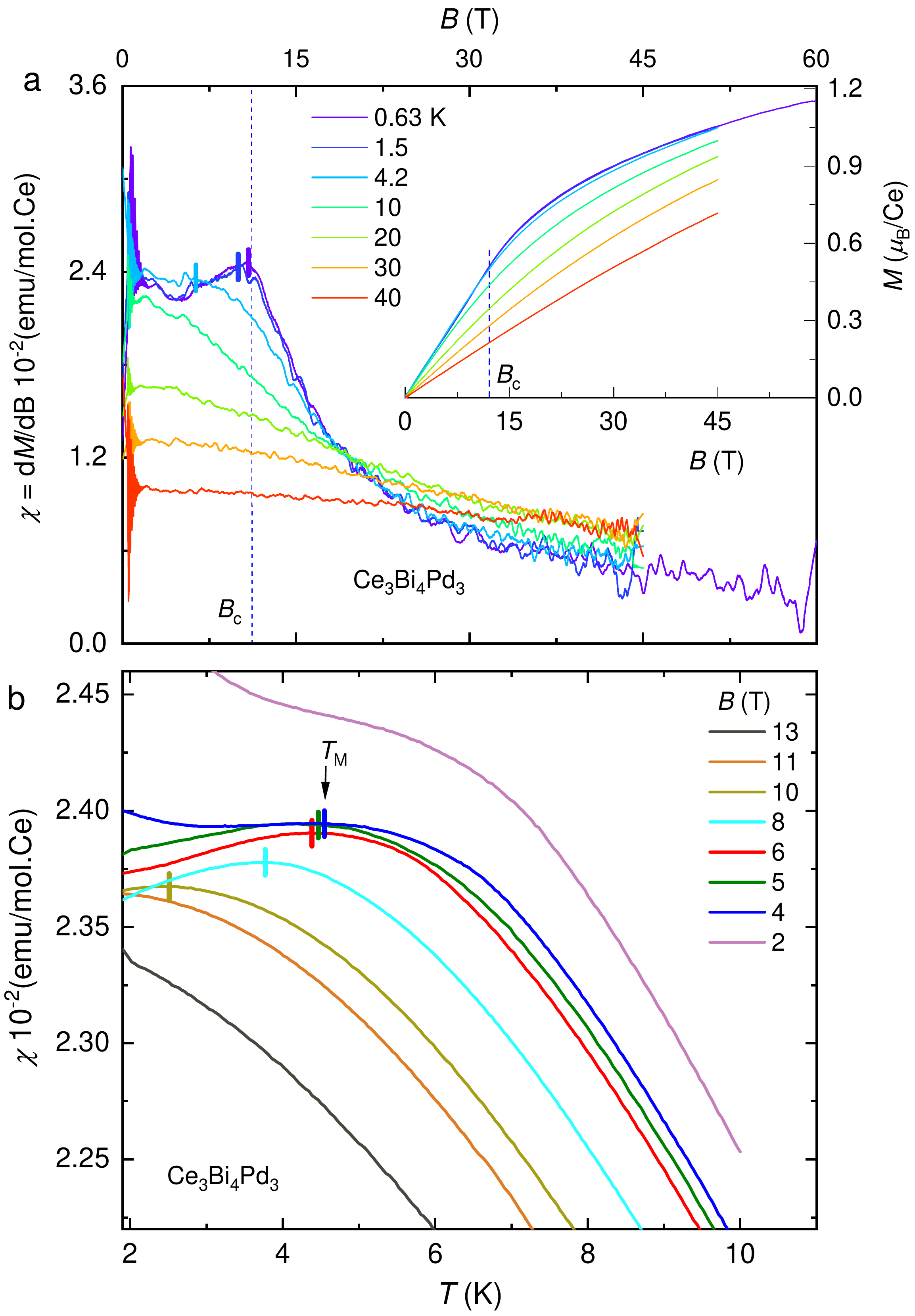}
\caption{{\bf a}, Magnetic susceptibility $\chi=\partial M/\partial B$ versus $B$ of Ce$_{3}$Bi$_{4}$Pd$_3$ (for $B\|\langle100\rangle$) at different temperatures (as indicated). Vertical bars indicate the $T$-dependent maximum. The inset shows the magnetization $M$ versus $B$.
{\bf b}, $\chi$ versus $T$ at different static magnetic fields (as indicated), with vertical bars indicating the maximum. Note that at $B=$~2~T, a shoulder rather than a peak is observed due to a low $T$ contribution from magnetic impurities (or defects) at low $B$ and $T$. A similar low $T$ contribution is observed in Ce$_3$Bi$_4$Pt$_3$ in Fig.~\ref{RES}b, although, in that case, it is farther from $T_{\rm M}$.
}
\label{MAG}
\end{figure}

Thermodynamic evidence linking the quantum critical point to the breakdown of the Kondo coupling accompanying the closing of the gap is provided by measurements of magnetization and magnetic susceptibility in Fig.~\ref{MAG}. A strong argument for the reduction in Kondo screening is provided by the approach of the magnetization towards saturation in the inset to Fig.~\ref{MAG}a, which is indicative of a state in which the $f$-electron moments are to a large extent polarized and subject to reduced fluctuations. More directly, we find the insulator-to-metal transition at $B_{\rm c}$ to be accompanied by a low-temperature peak in the magnetic susceptibility $\chi$ in Fig.~\ref{MAG}a, which we trace back to the peak in $\chi$ at $T_{\rm M}$ along the temperature axis. On performing magnetic susceptibility measurements over a range of temperatures and magnetic fields, we construct a pseudo phase boundary from the locus of the peak that continuously connects $B_{\rm c}$ with $T_{\rm M}$ within the $B,T$ plane (see Fig.~\ref{diagram}), revealing that $B_{\rm c}=B(T_{\rm M}\rightarrow$~0). In the low magnetic field limit, the broadness of the peak is indicative of a crossover, yet it becomes increasingly sharp on increasing the magnetic field and reducing the temperature, in a manner analogous to that reported in Sr$_3$Ru$_2$O$_7$.\cite{grigera2001} 

\begin{figure}
\includegraphics[angle=0,width=0.9\hsize]{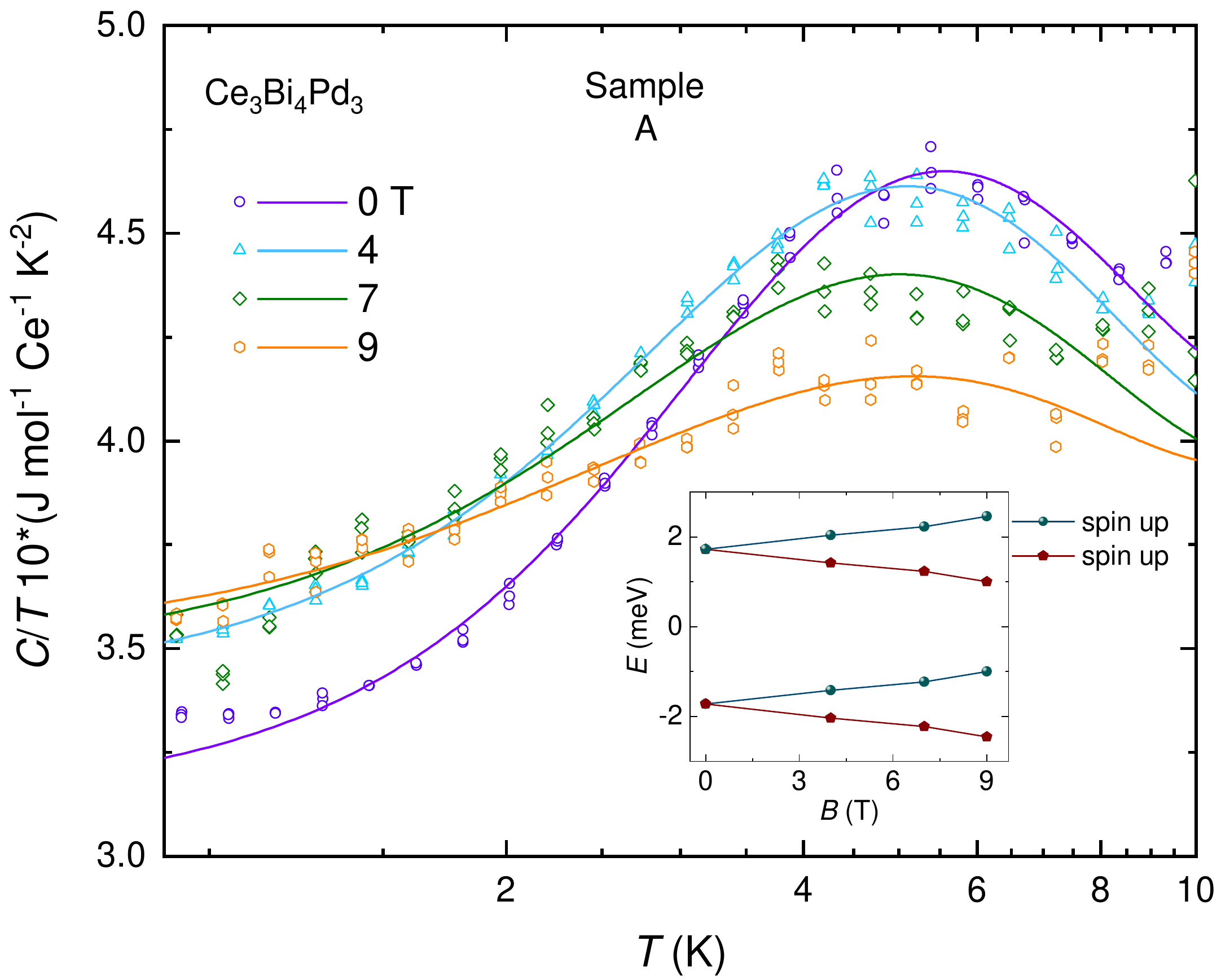}
\caption{{\bf a}, Low temperature heat capacity divided by temperature, $C/T$, plotted versus logarithmically-scaled $T$ (symbols). Solid lines represent fits in which the Kondo gap is modeled using the Schotte-Schotte model (see Appendix) at $B=0$ (purple line). On extending to fit to $C/T$ for $B=$~4, 7 and 9~T (light blue, green and yellow lines), only the Zeeman splitting $H=\pm\frac{1}{2}g_{\rm eff}\mu_{\rm B}B$ is allowed to vary. The inset shows the locations of the Lorentzians of width $\approx$~2~meV used in the Schotte-Schotte model (see Appendix). 
}
\label{heatcap}
\end{figure}

Finally, thermodynamic evidence for the magnetic field-induced closing of the Kondo gap is provided by the finding of a Schotte-Schotte anomaly in the heat capacity in Fig.~\ref{heatcap}a, which undergoes a continuous change in shape in an applied magnetic field. A Schotte-Schotte\cite{schotte1975,vandermeulen1991} anomaly occurs under circumstances in which a gap in the electronic density-of-states appears between peaks whose line-shapes have an approximate Lorentzian form. Fits (see Fig.~\ref{heatcap}a) indicate that the magnetic field-induced evolution of the Schotte-Schotte anomaly can be understood solely as the effect of Zeeman splitting on the gap, where the splitting energy is $H=\pm\frac{1}{2}g_{\rm eff}\mu_{\rm B}B$ and $g_{\rm eff}\approx$~2.9~$\pm$~0.3 is an effective {\it g}-factor for pseudospins of $\pm\frac{1}{2}$. On equating $2H=\Delta\approx k_{\rm B}T_{\rm K}$ at the critical magnetic field $B_{\rm c}$, we obtain $\Delta=$~1.8~$\pm$~0.2, which is therefore in excellent agreement with $\Delta$ estimated from Hall effect measurements (see Fig.~\ref{MRHall}b).

\section{discussion}

The signatures of quantum criticality in Figs.~\ref{diagram}, \ref{MRHall}, \ref{GAP} and \ref{MAG}, and their association with an insulator-to-metal transition, are strikingly similar to those recently identified in SmB$_6$ as a function of pressure.\cite{zhou2017,gabani2003}  In SmB$_6$, the entry into the metallic phase with increasing pressure can be qualitatively understood in terms of increased tendency for magnetism accompanying a continuous increase in the valence towards a trivalent value.\cite{cooley1995,gabani2003,barla2005,derr2006,zhou2017} In Ce$_{3}$Bi$_{4}$Pd$_3$, the polarization of the $f$-electron moments by a magnetic field causes the magnetic field to behave like an effective negative pressure; the lattice volume invariably increases under a magnetic field in Ce compounds,\cite{kaiser1988,zieglowski1986} causing a reduction in the valence towards a trivalent value. The relative ease by which $f$-electron moments are polarized by a magnetic field in Ce$_{3}$Bi$_{4}$Pd$_3$ (see inset to Fig.~\ref{MAG}a) compared to Ce$_{3}$Bi$_{4}$Pt$_3$\cite{modler1999} suggests that the smaller Kondo temperature and gap in the former is a likely consequence of it lying closer to integer valence than the latter.

The observed peak in the magnetic susceptibility at $B_{\rm c}$ has been predicted,\cite{oliveira1999} and suggests that metamagnetism,\cite{aoki2013,edwards1997}  such as that associated with the end point at the end of a line of first order transitions at or near $T=0$, provides one possible route for achieving quantum criticality.\cite{grigera2001,oliveira1999} An alternative possibility, given the suppression of the Kondo coupling in strong magnetic fields, is the sudden collapse of the Kondo coupling to a significantly smaller value at $B_{\rm c}$\cite{si2001} -- in effect causing the $f$-electrons to behave in a more localized fashion for $B>B_{\rm c}$. In both cases, the reduction in electronic correlations beyond $B_{\rm c}$ is expected to result in a gradual reduction in the effective mass $m^\ast$,\cite{edwards1997,aoki1993} or equivalently the $A$ coefficient,\cite{daou2006,grigera2001} with increasing magnetic field. The reduction in $A$ with increasing field in Ce$_{3}$Bi$_{4}$Pd$_3$ is indeed similar to that previously observed in $d$- and $f$-electron metamagnets (for example, CeRu$_2$Si$_2$).\cite{daou2006,grigera2001} 

The contiguity between a vanishing carrier density for $B<B_{\rm c}$ and suppressed Kondo coupling for $B>B_{\rm c}$ makes the insulator-to-metal transition in Kondo insulators somewhat unique compared to other $f$-electron systems. A crucial question, given the polarization of the $f$-electrons in a sufficiently strong magnetic field, concerns whether, at any point, the $f$-electrons contribute to the Fermi surface volume of Ce$_{3}$Bi$_{4}$Pd$_3$. 
For $B\gtrsim B_{\rm c}$, the Zeeman splitting of the hybridized bands is predicted to give rise to a semimetallic state composed of hybridized conduction and $f$-electron states.\cite{oliveira1999} The collapse of the Hall resistivity to a small value for $B\gtrsim B_{\rm c}$ is indeed consistent with a semimetal, implying that a partial participation of $f$-electrons in the Fermi surface will likely contribute to the enhancement of $A$ and the collapse of $T_{\rm FL}$ in the immediate vicinity of $B_{\rm c}$.\cite{edwards1997}  

Finally, there is the question of whether strong electronic correlations contribute to the growth of the electronic contribution to the heat capacity in the vicinity of the field-induced transitions into the metallic states of the Kondo insulators Ce$_3$Bi$_4$Pt$_3$ and YbB$_{12}$,\cite{jaime2000,terashima2018} or to the temperature-dependence of the quantum oscillation amplitudes within the insulating phases of SmB$_6$ and YbB$_{12}$.\cite{tan1,xiang2018} Our findings suggest, at the very least, that a growth of electronic interactions at the expense of a declining hybridization needs to be taken into consideration in future more complete models of the insulating state in strong magnetic fields. A further possibility is that the Kondo insulator gradually transforms into an excitonic insulator\cite{knolle2017} in advance of metallization.

This high magnetic field work was supported by the US Department of Energy `Science of 100 tesla' BES program. MKC acknowledges funding from the Los Alamos LDRD program, while SKK acknowledges support of the LANL Directors Postdoctoral Funding LDRD program. Sample characterization at Los Alamos National Laboratory was performed under the auspices of the U.S. Department of Energy, Office of Basic Energy Sciences, Division of Materials Sciences and Engineering, `Quantum Fluctuations in Narrow Band Systems' project. The high magnetic field facilities used in this work are supported by the National Science Foundation, the State of Florida and the Department of Energy. We acknowledge helpful discussions with Mucio Continentino, Peter Riseborough, Qimiao Si and Peter W\"{o}lfle. 

\section{Appendix}

\subsection{Sample preparation}
Ce$_3$Bi$_4$Pd$_3$ crystals are grown by the Bi-flux technique with starting composition Ce:Pd:Bi = 1:1:1.8. The reagents are put in an alumina crucible placed inside an evacuated quartz tube. The quartz tube is then heated to 1050$^\circ$C at 100$^\circ$C/h and is kept
there for 8 hours. The solution is then cooled down to
450$^\circ$C at 2$^\circ$C/h. The excess of Bi is removed
by spinning the tube in a centrifuge. The crystallographic structure is verified by single-crystal diffraction at room temperature using Mo radiation in a Bruker D8 Venture diffractometer. Bi-Pt superconducting binary phases are present as impurity phases.

\subsection{Electrical transport experimental details}
\begin{figure}[!!!!!!!htbp]
\centering 
\includegraphics*[width=.45\textwidth]{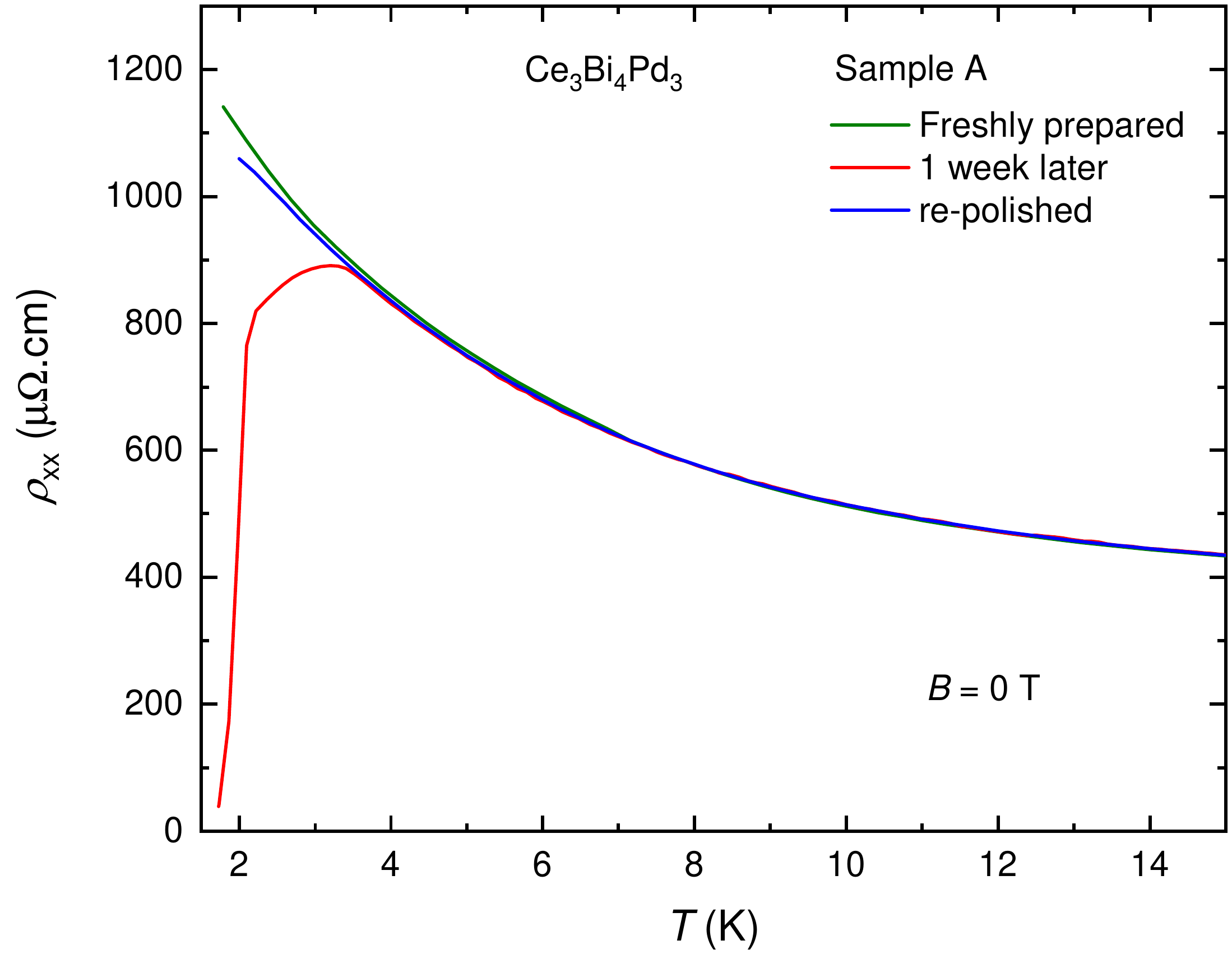}
\caption{The electrical resistivity of Ce$_{3}$Bi$_{4}$Pd$_3$ sample A, after being freshly prepared (green curve), thermally cycled (red curve) and re-polished to remove surface contamination (blue curve).}
\label{resistivity}
\end{figure}

Samples of  Ce$_{3}$Bi$_{4}$Pd$_3$ are cut into rectangular shapes and polished for electrical transport measurements. All resistivity measurements, including those in static and pulsed magnetic fields are performed using the four wire technique. Whereas freshly polished samples (green line in Fig.~\ref{resistivity}) exhibit insulating behavior down to $\approx$~2~K, thermally cycled samples exhibit additional superconductivity (red line in Fig.~\ref{resistivity}). After re-polishing, most of the additional superconductivity is removed (blue line in Fig.~\ref{resistivity}). 

\begin{figure}[!!!!!!!htbp]
\centering 
\includegraphics*[width=.45\textwidth]{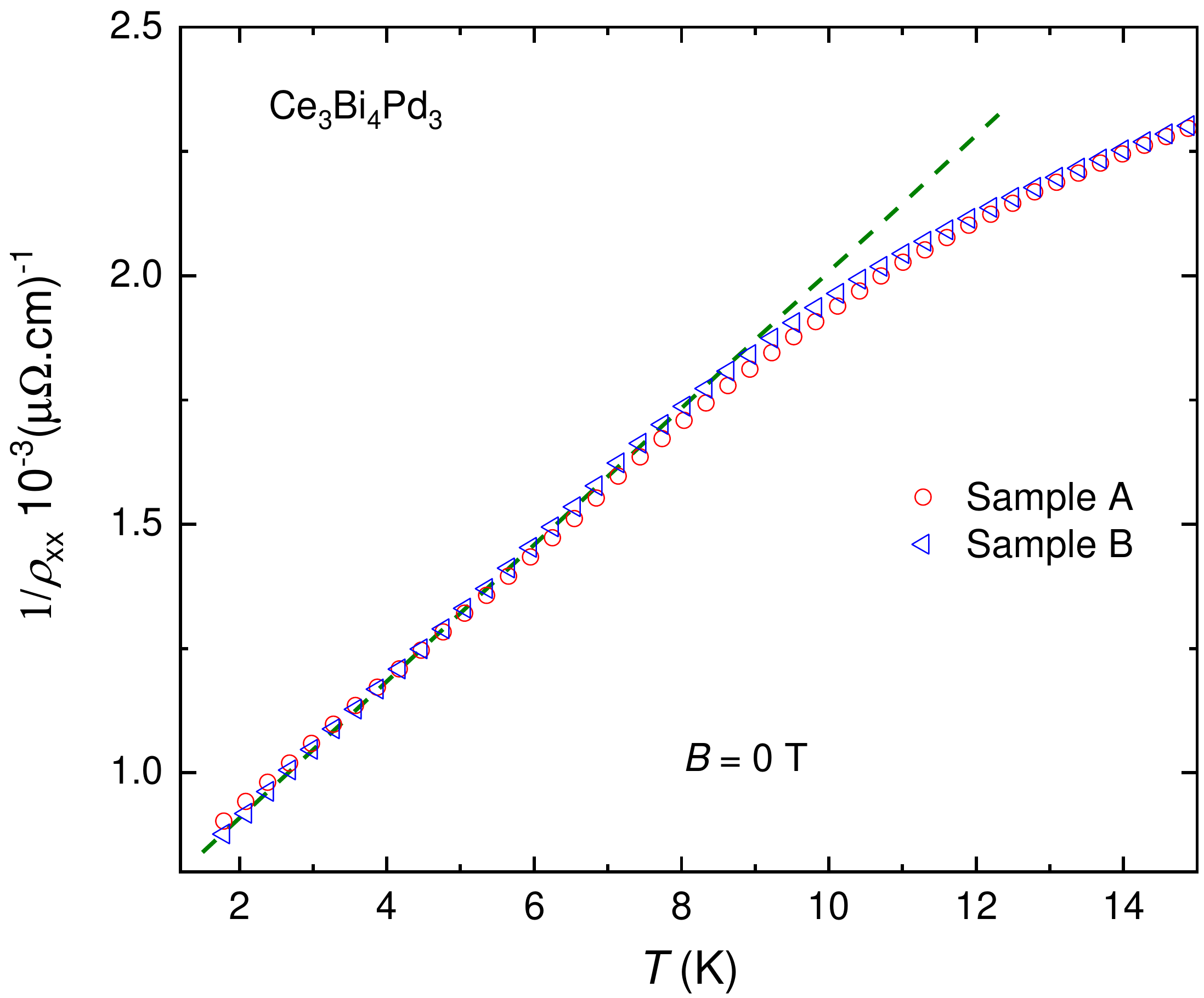}
\caption{Low temperature reciprocal resistivity of Ce$_{3}$Bi$_{4}$Pd$_3$ samples A and B, indicating a linear-in-$T$ component.}
\label{linear}
\end{figure}

Despite the opening of a gap in the energy spectrum, Kondo insulators are prone to in-gap states that can reside either at the surface or the bulk and generally cause the electrical resistivity to acquire a finite value in the limit $T\rightarrow$~0. In a similar manner to SmB$_6$,\cite{harrison2018} Ce$_3$Bi$_4$Pd$_3$ is found to have an inverse resistivity $\rho_{xx}^{-1}$ that varies linearly in $T$ at low temperatures (see Fig.~\ref{linear}), suggesting that scattering from defects may be responsible for the departure from activated behavior at the lowest temperatures (at least down to $\approx$~2~K).\cite{harrison2018} 

\subsection{Magnetic susceptibility experimental details}
Magnetic susceptibility measurements are performed using a Quantum Design vibrating sample magnetometer.
Unscreened rare earth impurities are known to give rise to an upturn in the susceptibility of Kondo insulators at low temperatures.\cite{menth1992,aeppli1992,maple1971,kasaya1985,nickerson1971} Owing to the low energy scale associated with the Kondo gap in Ce$_{3}$Bi$_{4}$Pd$_3$, a steady magnetic field of $B\gtrsim$~4~T is  required to uncover the maximum in the susceptibility. Only a shoulder is observed at lower magnetic fields. In a similar manner to Ce$_3$Bi$_4$Pt$_3$,\cite{modler1999} the magnetic susceptibility of Ce$_3$Bi$_4$Pd$_3$ continues to remain large as the magnetic field is increased, which is suggestive of a band Van Vleck paramagnetism effect.\cite{riseborough2000}

\subsection{Heat capacity experimental and modeling details}
\begin{figure}
\includegraphics[angle=0,width=0.9\hsize]{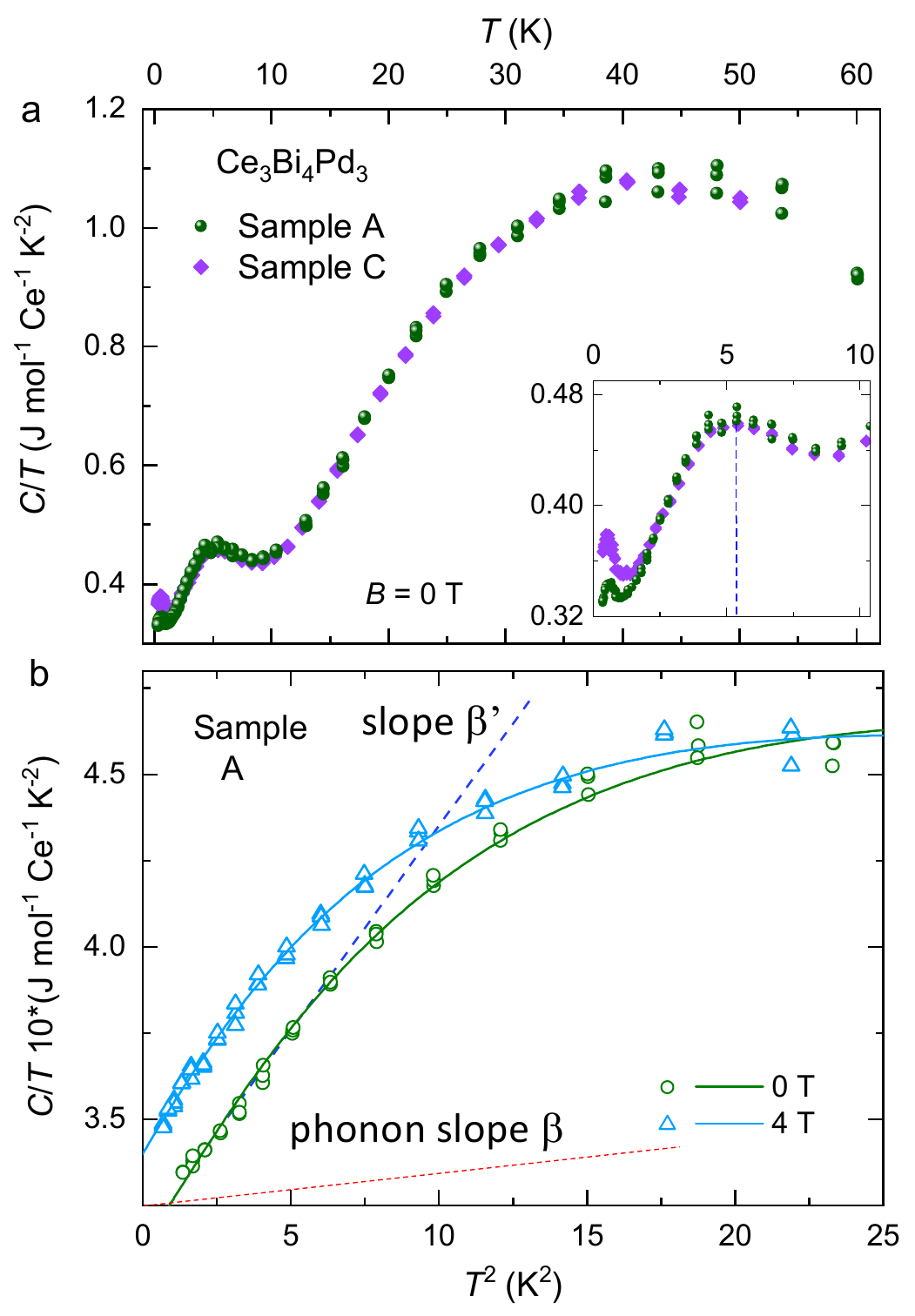}
\caption{{\bf a}, Heat capacity divided by temperature $C/T$ plotted versus $T$ for two samples. The inset shows an expanded view of the low temperature region. {\bf b}, Low temperature portion of $C/T$ plotted versus $T^2$, showing an approximate $T^2$ contribution at low $T$ on top of that $\beta T^2$ due to phonons (the assumed phonon contribution is shown in red).
}
\label{heatcap2}
\end{figure}

Heat capacity measurements on Ce$_3$Bi$_4$Pd$_3$, made using a Quantum Design Physical Properties Measuring System, reveal a peak in the overall heat capacity centered at $\approx$~5~K (see Fig.~\ref{heatcap2}a). In Ce$_3$Bi$_4$Pt$_3$, by contrast, a peak is observed in the heat capacity centered on $\approx$~50~K.\cite{jaime2000,riseborough2000} The heat capacity is found to be reproducible in different samples for $T\gtrsim$~1.5~K. Features that appear below $\approx$~1.5~K are sample-dependent, suggesting them to be of extrinsic origin.

\begin{figure}
\includegraphics[angle=0,width=0.9\hsize]{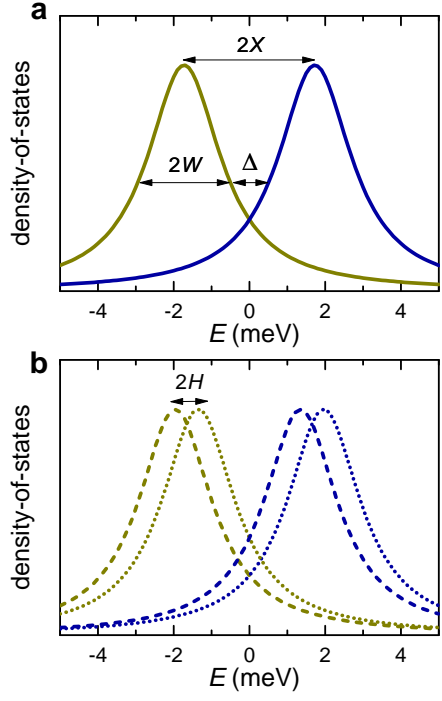}
\caption{{\bf a}, Density-of-states of the upper (royal blue) and lower (dark yellow) bands in the form of Lorentzians used in the Schotte-Schotte model.\cite{schotte1975} In the model, we assume the gap be be symmetric. {\bf b}, The same bands split by a Zeeman splitting energy $2H$.
}
\label{schotte}
\end{figure}

The Schotte-Schotte anomaly is modeled here using an electronic density of states of the form 
\begin{equation}\label{lorentzian}
D(\varepsilon)=D_0\sum_{j=\pm1,\sigma=\pm\frac{1}{2}}\frac{W}{\pi(W^2+(\varepsilon+jX+\sigma g_{\rm eff}\mu_{\rm B}B)^2)},
\end{equation}
with the free energy given by $F=\int^\infty_{-\infty}D(\varepsilon)\ln[1+\exp(-\varepsilon/k_{\rm B}T)]{\rm d}\varepsilon$ and the heat capacity given by $C_p\approx C_v=T\partial^2TF/\partial T^2|_v$, which we calculate numerically\cite{silhanek2005} and fit to the experimental data over the temperature range 1.5~$\lesssim T\lesssim$~10~K. Here, $\pm X$ refers to the peaks in the electronic density-of-states above and below the chemical potential (see Fig.~\ref{schotte}a), which are responsible for producing the Kondo gap at $E=0$, while $W$ refers to their broadening, which produces bulk in-gap states.
Meanwhile $g_{\rm eff}$ is an effective {\it g}-factor describing the Zeeman splitting (see Fig.~\ref{schotte}b) of the density-of-states under a magnetic field.

During fitting, we assume a phonon contribution $C_{\rm ph}/T=\beta T^2$ where $\beta=$~1~mJmol$^{-1}$Ce$^{-1}$K$^{-4}$. On fitting to $C/T$ under a magnetic field, $W$ and $X$ are held at their $B=0$ values while $|H|$ is allowed to vary as a fitting parameter. The magnitude of $H$ suggests that $g_{\rm eff}\approx$~2.9. According to the fit, the spacing of the upper and lower Lorentzian-shaped (or broadened) bands is $2X\approx$~3.45~meV while their widths are $2W\approx$~2.25~meV, implying an interband gap of order $\Delta\approx2X-2W\sim$~1.2~meV. This is of comparable order to $\Delta$ estimated from Hall effect measurements (see Fig.~\ref{MRHall}a) and that obtained by equating $2H=\Delta$.  

A distinguishing feature of the Schotte-Schotte anomaly that appears to be present in the experimental data (see Fig.~\ref{heatcap} and Ref.~\cite{dzsaber2017}) is an electronic contribution to the heat capacity of the form 
\begin{equation}\label{cubic}
C_{\rm el}/T\approx\gamma_0+(\beta^\prime-\beta)T^2
\end{equation}
(see Fig.~\ref{heatcap2}b) in the limit $T\rightarrow0$, where $(\beta^\prime-\beta)T^2$ exceeds the contribution $\beta T^2$ from phonons and where $\gamma_0$ is the Sommerfeld coefficient characterizing in-gap states (that also depends on the strength of the magnetic field).\cite{silhanek2005}  Figure~\ref{heatcap2}b shows that Ce$_3$Bi$_4$Pd$_3$ exhibits the form given by Equation (\ref{cubic}) in the limit $T\rightarrow0$, with $\gamma_0$ increasing by $\approx$~30~mJmol$^{-1}$K$^{-2}$ on increasing the magnetic field from 0 to 4~T. A comparable increase was reported in Ce$_3$Bi$_4$Pt$_3$ on increasing the magnetic field from 0 to 40~T.\cite{jaime2000}


\subsection{Anomalous Hall effect estimate}
We estimate the anomalous Hall effect contribution from skew scattering within the high magnetic field metallic phase using $\rho_{xy}^{\rm A}=\gamma\tilde{\chi}\rho_{xx}$,\cite{fert1987} where $\gamma\sim$~0.08~KT$^{-1}$ is a typical value of the coefficient for Ce compounds.\cite{fert1987} At $B=0$, the reduced susceptibility is $\tilde{\chi}\approx\chi^{\rm peak}_{\rm 5~T}/C\approx$~0.03~K$^{-1}$, but this falls to $\approx$~0.014~K$^{-1}$ at 60~T. Using $\rho_{xx}\approx$~160~$\mu\Omega$cm at 60~T, we obtain $\rho_{xy}^{\rm A}\sim$~1.5~$\times$~10$^{-8}$~$\Omega$m at 60~T, which is $\sim$~10\% of $\rho_{xy}$.

\end{document}